\documentclass[sigconf]{acmart}
\AtBeginDocument{%
  \providecommand\BibTeX{{%
    \normalfont B\kern-0.5em{\scshape i\kern-0.25em b}\kern-0.8em\TeX}}}

\renewcommand\footnotetextcopyrightpermission[1]{}
\usepackage{amsmath,amsfonts}
\usepackage{algorithmic}
\usepackage{algorithm}
\usepackage{array}
\usepackage[caption=false,font=normalsize,labelfont=sf,textfont=sf]{subfig}
\usepackage{textcomp}
\usepackage{stfloats}
\usepackage{url}
\usepackage{verbatim}
\usepackage{graphicx}
\usepackage{xcolor,soul}
\usepackage{physics}
\setcopyright{none}
\begin{document}

\title{Optimization of Quantum Read-Only Memory Circuits}

\author{Koustubh Phalak}
\email{krp5448@psu.edu}
\affiliation{
 \institution{Pennsylvania State University}
 \state{PA}
 \country{USA}
}
\author{Mahabubul Alam}
\email{mxa890@psu.edu}
\affiliation{%
  \institution{Pennsylvania State University}
  \state{PA}
  \country{USA}
}

\author{Abdullah Ash-Saki}
\email{axs1251@psu.edu}
\affiliation{%
  \institution{Pennsylvania State University}
  \state{PA}
  \country{USA}
}

\author{Rasit Onur Topaloglu}
\email{rasit@us.ibm.com}
\affiliation{%
  \institution{IBM New York}
  \state{NY}
  \country{USA}
}

\author{Swaroop Ghosh}
\email{szg212@psu.edu}
\affiliation{%
  \institution{Pennsylvania State University}
  \state{PA}
  \country{USA}
}

\begin{abstract}
  Quantum computing is a rapidly expanding field with applications ranging from optimization all the way to complex machine learning tasks. Quantum memories, while lacking in practical quantum computers, have the potential to %be key circuit primitives that promise to 
  bring quantum advantage. %in solving the problem at hand. 
  In quantum machine learning applications for example, a quantum memory can simplify the data loading process and potentially accelerate the learning task. Quantum memory can also store intermediate quantum state of qubits that can be reused for computation. However, the depth, gate count and compilation time of quantum memories such as, Quantum Read Only Memory (QROM) scale exponentially with the number of address lines making them impractical in state-of-the-art Noisy Intermediate-Scale Quantum (NISQ) computers beyond 4-bit addresses. %A QROM with short depth and low gate count on the other hand could improve the fidelity of a circuit. 
  In this paper, we propose techniques such as, predecoding logic and qubit reset to reduce the depth and gate count of QROM circuits to target wider address ranges such as, 8-bits. The proposed approach reduces the number of gates and depth count by at least 2X compared to the naive implementation at only 36\% qubit overhead. A reduction in circuit depth and gate count as high as 75X and compilation time by 85X at the cost of a maximum of 2.28X qubit overhead is observed. Experimentally, the fidelity with the proposed predecoding circuit compared to existing optimization approach is also higher (as much as 73\% compared to 40.8\%) under reduced error rates.
%   \hlkp{We also compare the results with even further optimized circuit and the originally proposed circuit.}  The next generation NISQ computers are projected to contain hundreds of qubits but expected to suffer from low Quantum Volume (QV) (note, QV is a measure of computing power of quantum computer under various quantum noise). Therefore, our approach of sacrificing the qubits in favor of shallow and low gate count QROM circuit is appropriate for NISQ computers.
\end{abstract}





\maketitle
\pagestyle{plain}

\section{Introduction}
Memory acts as a bridge between the processor and the storage element, where the frequently accessed data is stored in a volatile or non-volatile manner. This saves extra time to fetch the data from the slow storage element. In the quantum domain, all qubits are initialized to a value such as $\ket{0}$. %which is defined by a given quantum computing hardware. 
Therefore, data needs to be loaded in the circuit first before performing a computation. The data loading can be performed via various embedding methods such as amplitude embedding, angle embedding, and hybrid embedding \cite{schuld2021effect}. Each of these methods bring their own set of benefits and challenges. Amplitude embedding, on one hand, can accommodate $2^n$ data into $n$ qubits at higher circuit depth and gate count degrading the fidelity of computation. Angle embedding, on the other hand, encodes the data as the rotation angle along X/Y/Z axis of a single qubit rotation gate. Therefore, $n$ data points can be loaded onto $n$ qubits relieving the qubit count requirement. One can encode more than one data in a qubit by cascading rotation gates \cite{perez2020data} at the cost of increased circuit depth. In quantum machine learning (QML) applications, data loading present significant training time overhead since the classical dataset needs to be uploaded in quantum domain iteratively and the output sampled in classical domain to determine the gradient and optimize the parameters. Efficient data encoding is an active area of research.

%Quantum memory can address the problem of data loading since the data can be read in the quantum circuit when needed. 
In QML applications, quantum memory can simplify the training since the data can be loaded and processed within the quantum circuit without converting to the classical domain. Quantum memory can also store intermediate quantum states during computation to reclaim the qubits. Various circuit-based Quantum Random Access Memory (QRAM) and Quantum Read Only Memory (QROM) \cite{park2019circuit,babbush2018encoding} circuits using Noisy Intermediate-Scale Quantum (NISQ) computers have been proposed. However, they incur exponential circuit depths and gate counts with the number of address lines degrading the fidelity of the computation. This renders the quantum memory slow and useless.  Optimizations techniques are warranted to address this challenge.

Our work is related to the QROM implementation \cite{babbush2018encoding} where a common sub-circuit called 'unary iteration' is used for controlling the data to be sent on the data lines. However, this circuit uses multi-controlled not gates which decomposes into large number of basis gates increasing the depth as the number of control lines (which is a function of address width in QROM) grow. Moreover, the number of such multi-controlled not gates grows exponentially with increasing address lines due to the structure of the QROM circuit. A sawtooth circuit is implemented for optimization where multi-controlled not gates are broken down into large number of toffoli gates with the help of ancilla qubits added between address qubits. However, the count of multi-controlled not gates even in the optimized circuit once again is exponential. Thus, the increase of overall gate count is exponential with increasing number of address lines  %while this optimization is helpful to bring down the circuit depth and the gate count while compared to that of the naive implementation of the circuit, 
leaving room for more robust optimization. In this paper, we propose optimizations of the unary iteration sub-circuit to reduce the compilation time, gate count, and circuit depth of the QROM circuit for faster access.

In the remaining of the paper, Section 2 presents the relevant background details and related works. Section 3 describes the optimizations performed on the QROM circuit. Section 4 compares the results with the naive QROM implementation. The limitations are also discussed. Section 5 presents a general discussion. Finally, Section 6 concludes the paper.

\section{Background and Related Works}
\subsection{Quantum computing fundamentals}

\noindent \textbf{Qubits:} Quantum bits or qubits are the fundamental units of a quantum computer. While classical bits can have two possible values of zero or one, qubits have quantum states denoted using the ket notation $\ket{\psi}$. This state can carry the probability $a^2$ of it being $\ket{0}$, and the probability $b^2$, of it being $\ket{1}$, where $a$ and $b$ are the complex numbers with ($a^2+b^2=1$). Therefore, qubit can exist in both the states simultaneously. Qubits are also represented in matrix form. For example, $\ket{0}$ is denoted as 
$\big[\begin{smallmatrix}
  1\\
  0
\end{smallmatrix}\big]$ 
and $\ket{1}$ is denoted as 
$\big[\begin{smallmatrix}
  0\\
  1
\end{smallmatrix}\big]$. 

\noindent \textbf{Qutrits:} Qutrits are a ternary version of qubits which can store states of three classical values instead of just zero and one. In the context of QRAM, a qutrit has left, right, and wait states (\cite{giovannetti2008quantum}). Qutrit can be realized in a trapped ion quantum computer with the different states as different energy levels of the trapped ion. Routing direction is determined based on current qutrit state.
% Depending on the current state of the qutrit, the qubits either set state of the current qutrit, or are routed accordingly to next set of qutrits which ultimately leads the qubit into the memory cell to be accessed in the QRAM.\\ 

\noindent \textbf{Quantum Gates:} A quantum circuit has quantum gates, which perform operations on qubits and change their state. Quantum gates can be represented as a unitary matrix. 
%The different types of gates are categorized based on the number of qubits they operate on. 
\begin{figure}[t]
    \centering
    \includegraphics[width=0.5\linewidth]{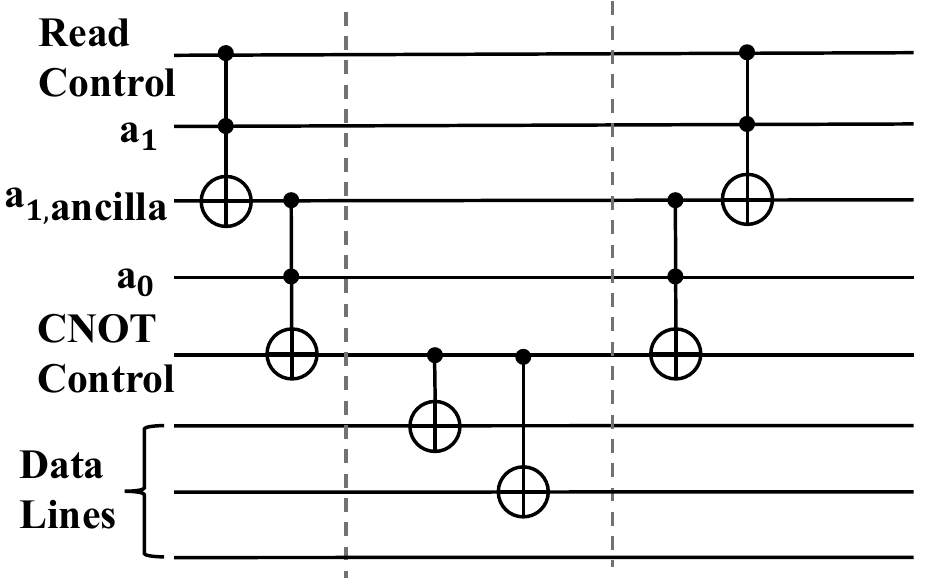}
    \caption{Sawtooth circuit implementation of QROM as proposed in \cite{babbush2018encoding}}
    \label{fig:sawtooth}
\end{figure}

The most frequently used categories of gates are single qubit gates, which operate on one qubit, and two qubit gates, which operate on two qubits at once. Hadamard (H), Bit flip (X) and Rotation gate (RX, RY, RZ) are commonly used single qubit gates while Controlled Not (CNOT) is a commonly used two-qubit gate. More than two qubit gates exist as well such as, Contolled Swap, Peres, Toffoli, iToffoli, etc. Every quantum hardware has a set of associated basis gates which every complex gate is broken down into. For example, the current IBM backends use RZ, ID, X, SX, and CNOT gates as the basis gates. When a quantum circuit is sent to the quantum hardware, the complex multi-qubit gates are decomposed into these basis gates. %This process is called decomposition. 
The gate count and the overall depth of the decomposed quantum circuit depends on the complexity of the multi-qubit gates.\\

\noindent \textbf{Quantum Errors:} One of the major challenges faced by modern NISQ computers is the wide range of errors. Readout error or measurement error occurs while measuring a qubit from the quantum to classical state. Gate error occurs when a quantum gate gives a wrong computational output due to environmental noise and errors in microwave/laser pulses. Decoherence error happens in deep quantum circuit due to environmental factors such as, temperature. When two gate operations are performed in parallel on neighboring qubits they interfere with each other to corrupt the qubit state. This is called crosstalk error. Various errors accumulate with circuit size. % and major sources of error. That is why, it is important to have 
Therefore, shallow and small quantum circuits are preferred for noisy quantum computers. \\

\subsection{QROM} 
\begin{figure}[t]
    \centering
    \includegraphics[width=\linewidth]{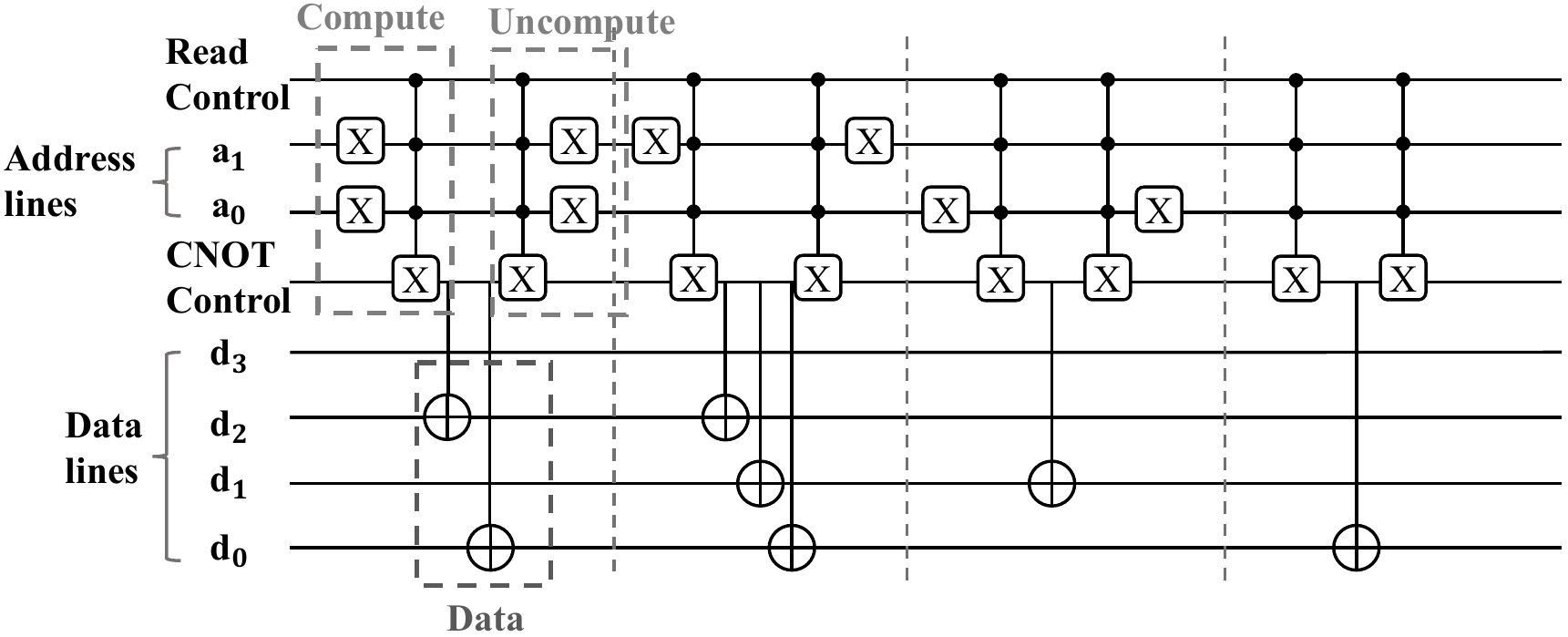}
    \caption{Naive implementation of QROM circuit with two address lines. In this example, the address and data values are as follows: QROM[0] = 5, QROM[1] = 7, QROM[2] = 2, QROM[3] = 1.}
    \label{fig:naive_qrom}
\end{figure}

In Read-Only Memory (ROM), the user can only read the data but cannot write into it. The QROM circuit has a read control signal, address lines, an extra ancilla qubit for CNOT control of data, and data lines. The read control signal provides the read signal to the quantum memory. The user can read the data only if the read signal is in state $\ket{1}$.
%, otherwise the user cannot read the data. 
When the read signal is active, the user will provide valid input on the address lines and get the output on the data lines. The data output depends on the Multi-controlled CNOT gate that gets activated based on the value of address lines. Initially, all the data lines are in $\ket{0}$ state. The naive implementation of the QROM circuit (Fig. \ref{fig:naive_qrom}) contains two address lines and four data lines. To optimize this naive implementation, \cite{babbush2018encoding} proposed the sawtooth circuit of the unary iteration. In the sawtooth circuit, ancilla qubits are inserted between the address qubits, and the multi-controlled not gate is broken down into toffoli gates. Assuming two address lines, there will be two toffoli gates. The first toffoli gate will be controlled by read control and the MSB $a_1$ with target at $a_{1,ancilla}$, and the second toffoli gate will have be controlled by $a_{ancilla}$ and $a_0$ with target at the CNOT control line. This circuit is shown for a single datapoint in Fig. \ref{fig:sawtooth}. In this approach, $O(n)$ extra qubits are required for n address lines.
%We later also present the proposed optimized version of this naive implementation shown in Fig. \ref{fig:optimized_qrom}, which uses the pre-decoding scheme that will be explained in more detail in Section 3. 

\subsection{Related Work}
Our work is closely related to \cite{babbush2018encoding} that primarily focuses on usage of quantum computing for quantum physics and quantum chemistry. One of the techniques used in their circuits is `unary iteration', which is a set of control qubit lines to perform control operations. This unary iteration has been used in QROM circuit as one of it's applications. However, there is little optimization performed on the unary iteration circuit resulting in a deep and large gate count overhead over the QROM circuit (details in Section 3). %In this paper, we show that optimization of the unary iteration circuit can drastically decrease both gate count and depth of the QROM circuit.\\

Other related works involve QRAM development. The quantum version of bifurcation graph-based RAM is proposed in \cite{giovannetti2008quantum} which utilizes qutrits to route the qubits to appropriate memory cells. The bifurcation graph-based RAM can be represented as a full binary tree in which the leaves represent the memory cells, and rest of the nodes are qutrits which assist in routing the qubits to the appropriate memory location. It has both exponential circuit width and circuit depth. Another work \cite{arunachalam2015robustness} analyzes the robustness of this QRAM architecture. Various possible architectures of QRAM are also proposed  \cite{giovannetti2008architectures}. A Flip-Flop QRAM architecture is proposed in \cite{park2019circuit} which store data into qubits in the form of superposition of states. The flip stage (which is the compute stage) loads each data, a register stage stores the data into the register qubit, and a flop stage performs uncomputation on the data lines. The application of FF-QRAM is also extended to continuous amplitudes in \cite{veras2020circuit}, which extend their application to loading continuous data instead of only discrete data.

Potentials applications for quantum memories are also studied e.g., usage of Raman quantum memory for optical quantum computing \cite{heshami2016quantum}. A detailed explanation on quantum cryptography with integration of quantum memories into quantum repeaters has been presented in \cite{pirandola2020advances}. An application of quantum memories in quantum communication has been mentioned in \cite{orieux2016recent}.

\begin{figure}[t]
    \centering
    \includegraphics[width=0.55\linewidth]{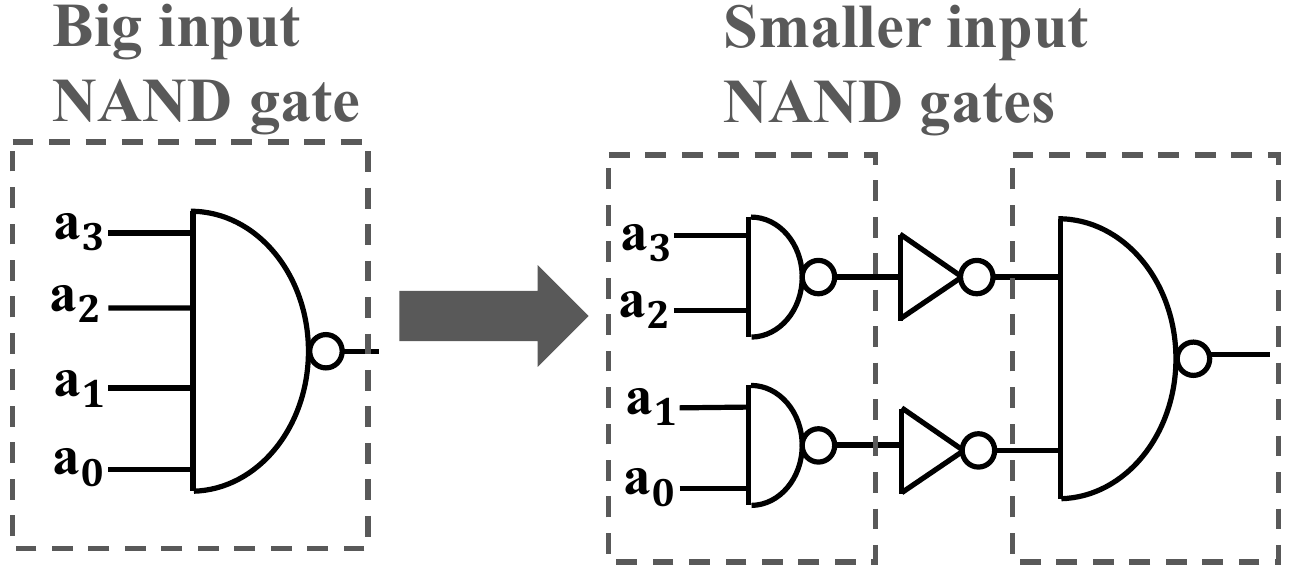}
    \caption{Classical pre-decoding used in implementing wide input NAND gates (4-input in this example) in memory array decoding logic using small fanin NAND gates. The outputs from first two NAND gates (located in midlogic area) are the pre-decoded signals that are provided to the final NAND gate (located in wordline driver area) for decoding.}
    \label{fig:classical_predecoding}
\end{figure}

\section{Pre-decoding in QROM circuits}
\subsection{Naive Implementation}
The naive implementation of the unary iteration sub-circuit of the QROM circuit \cite{babbush2018encoding} consists of a read control line, address lines, and a CNOT control line. For every data point, it consists of three stages: compute stage, data read stage, and uncompute stage as marked in three boxes respectively, in Fig. \ref{fig:naive_qrom}. In the compute stage, a multi-controlled not (MCX) gate is used where the controls are on the read control line and address lines, and the target is on the CNOT control line of the data lines. In general, $C^{n+1}X$ (controlled not gate having $n+1$ control signals) gates are required for control for $n$ address lines. Moreover, X gates are added prior to the MCX gates to flip address lines in $\ket{0}$ state to $\ket{1}$ state and activate all the controls of the MCX gate pertaining to that particular address only. For example, in Fig. \ref{fig:naive_qrom}, if $a_1 = \ket{0}$ and $a_0 = \ket{0}$, then the first set of X gates will flip the state of both the address lines to $\ket{1}$ state. Assuming that the read control line is also at $\ket{1}$ state, only the first MCX gate of the compute will be triggered. This will flip the CNOT control line to $\ket{1}$ state, and the data lines will read the data using the CNOTs in between the MCX gates. Finally, an uncompute stage is required to flip the CNOT control state back to $\ket{0}$ otherwise the CNOTs designated for other addresses will get triggered corrupting the original data.\\

\begin{figure*}[t]
\includegraphics[width=\linewidth]{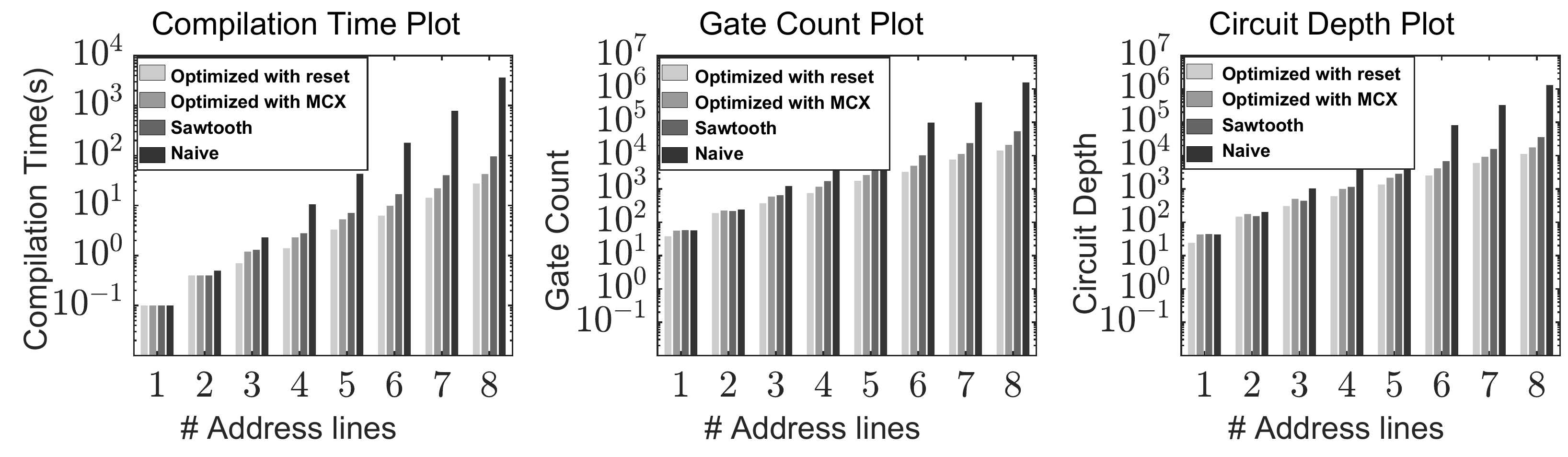}
\caption{Comparison of compilation time, circuit depth and gate count QROM circuit for both naive and optimized implementations with varying number of address lines. In the naive implementation, each performance metric value approximately increases $\sim4x$ for each extra address line. This is due to $4x$ increase in the number of MCX gates for each extra address line.}
\label{table:naive_results}
\end{figure*}

This naive structure of MCX gates works well when the number of address lines is small. However, as the number of address lines increases, so does the number of control lines required for the MCX gates which are broken down into basis gates during the compilation process. An MCX gate with $n+1$ controls takes at least 2X more number of basis gates than an MCX gate with $n$ controls for proper decomposition. Thus, the decomposition of the MCX gates increases the gate count, and consequently, the overall gate count and depth of the QROM circuit drastically increases as the number of address lines increase. Moreover, structure of the naive QROM circuit also leads to exponential number of such MCX gates. Suppose a QROM circuit has $n$ address lines which can store $2^n$ data. For each data, two MCX gates are required, one each for the compute and uncompute stages. Therefore, in total $2 * 2^n = 2^{n+1}$ MCX gates will be used. Also, let's assume that in the best case, an MCX gate can be decomposed into $O(n)$ Toffoli gates (\cite{linearMCXreduction}). Therefore, the total number of Toffoli gates after decomposition becomes $O(n)*2^{n+1} = O(n*2^n)$. Toffoli gates can be broken down into basis gates using $O(1)$ basis gates. Therefore, overall gate count of the circuit will be $O(1) * O(n*2^n) = O(n*2^n)$. Therefore, the increase in gate count is exponential with the address lines as shown in Fig. \ref{table:naive_results}.

\begin{figure}[t]
    \centering
    \includegraphics[width=0.5\linewidth]{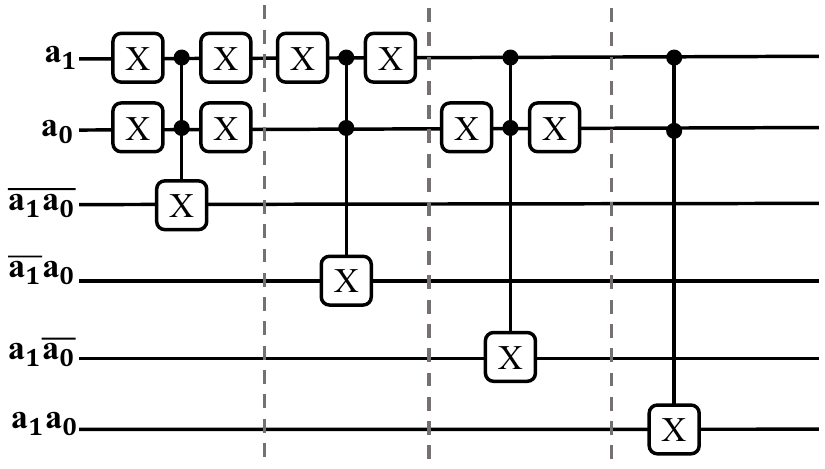}
    \caption{Pre-decoding of two address lines in QROM circuit to generate 4 signals $\overline{a_1a_0}$, $\overline{a_1}a_0$, $a_1\overline{a_0}$ and $a_1a_0$.}
    \label{fig:predecoding}
\end{figure}

\subsection{Proposed Pre-decoding Implementation}
In order to prevent this exponential increase of gate count, we propose pre-decoding of address lines. Similar to pre-decoding performed in classical memory, a subset of address lines are taken and all possible combinations of their signals are generated beforehand prior to providing them as input to the final decoder. For example, it is difficult to realize a 4 input NAND gate inside wordline driver due to large footprint. To overcome this issue, the NAND gate is broken down into 2 input NAND gates by performing pre-decoding. Two pre-decoded signals are generated using the two NAND gates at two pairs of address lines, and a final NAND gate is used with two NOT gates in between on these two pre-decoded address lines. This is shown in Fig. \ref{fig:classical_predecoding}. Therefore, design complexity is reduced at the cost of extra pre-decoded signals.

In QROM circuit, the pre-decoded signals are obtained on extra ancilla qubits before sending to the MCX gates. This reduces the number of control operations on the MCX gates, thereby shortening its decomposition. In general, for $m$ address lines pre-decoded together, the $m$ controls are reduced to just one single control. Moreover, multiple subsets of address lines can be pre-decoded separately for further reduction in gate count. Fig. \ref{fig:predecoding} shows the pre-decoding operation on two address lines $a_1$ and $a_0$ to generate $2^2=4$ pre-decoded signals. Therefore, 4 extra ancilla qubits are required. For pre-decoding of $m$ address lines, $2^m$ extra ancilla qubits are needed increasing the circuit width. However, this approach reduces the circuit depth and gate count improving the circuit performance under quantum errors. %Therefore, this will be beneficial overall for the QROM circuit. 
It should be noted though, that the increase in gate count will still be exponential, but it will not be as drastic due to the reduction in number of control signals of MCX gates.

\begin{figure}[t]
    \centering
    \includegraphics[width=\linewidth]{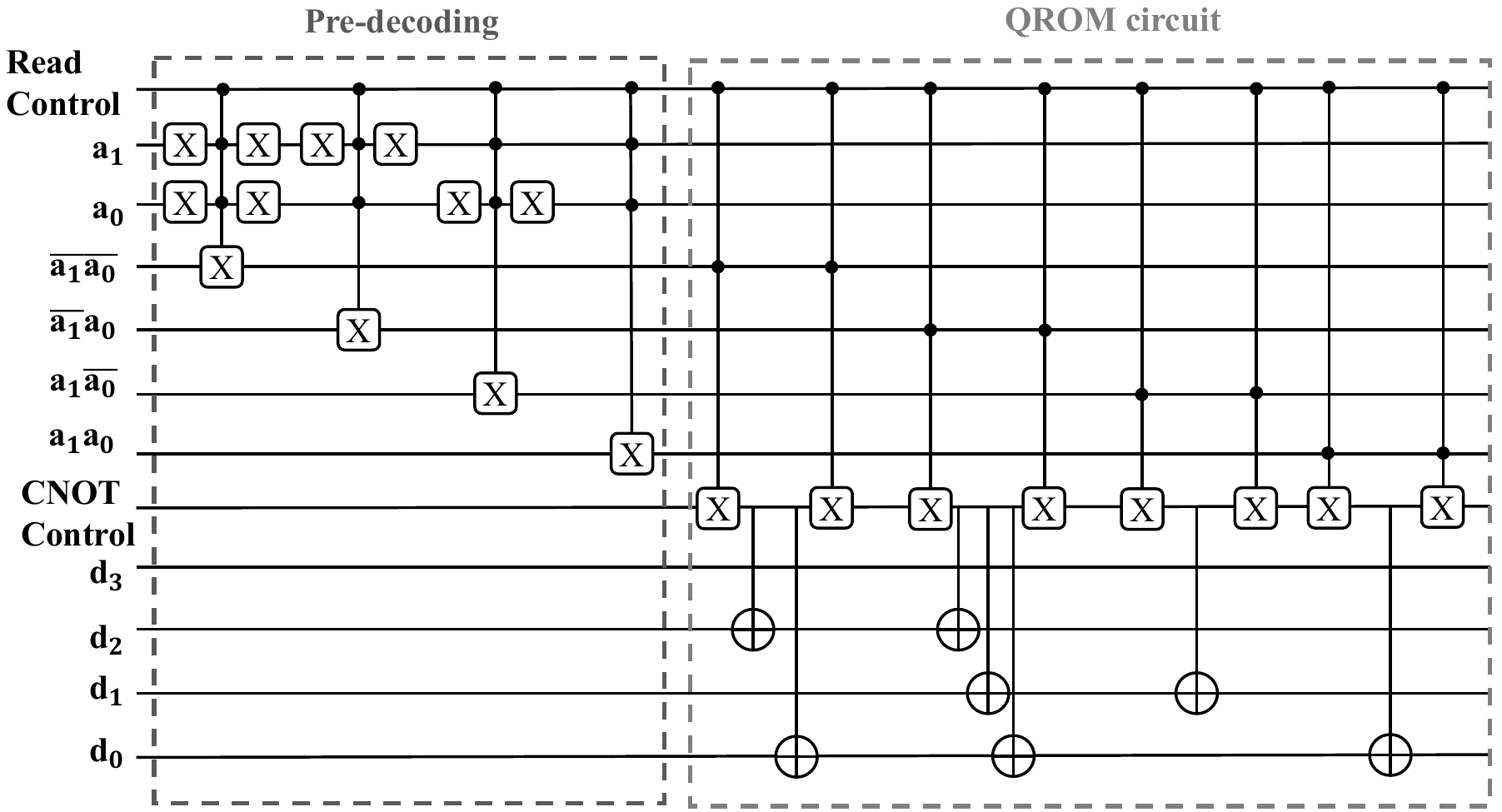}
    \caption{Optimized QROM circuit. The number of $C^{n+1}X = C^3X$ gates has reduced compared to the naive implementation.}
    \label{fig:optimized_qrom}
\end{figure}

\begin{figure}[t]
    \centering
    \includegraphics[width=\linewidth]{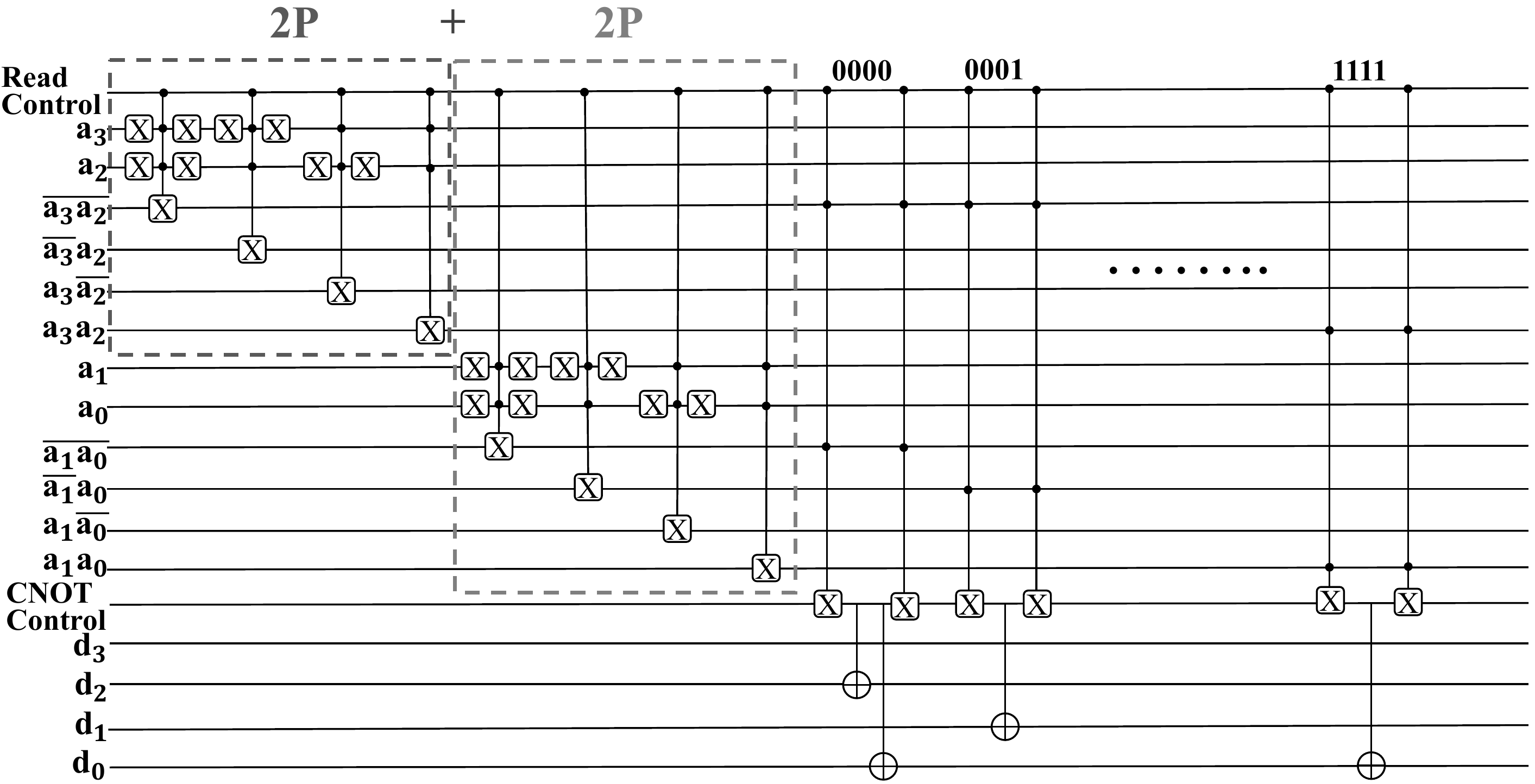}
    \caption{Pre-decoded QROM circuit with 4 address lines pre-decoded as 2(\textbf{P})+2(\textbf{P}).}
    \label{fig:2p_2p}
\end{figure}

Incorporating the pre-decoding scheme (e.g., Fig. \ref{fig:predecoding}) into the naive implementation (e.g., Fig. \ref{fig:naive_qrom}), we get the optimized version of the QROM circuit as shown in Fig. \ref{fig:optimized_qrom}. Comparing both the naive and optimized implementations, we note a reduction in the larger $C^3X$ gates. This is because some address signals were already pre-decoded. We can further reduce the number of multi-controlled not gates by replacing the MCX gates in the uncompute stage with a reset gate. A reset gate is a single qubit gate which resets the state of the qubit state back to $\ket{0}$ state. In the uncompute stage, the CNOT control line is required to be reverted back to $\ket{0}$ state. This is because if the state of the CNOT control line is not reset, unwanted CNOT gates corresponding to other address lines may get triggered and output wrong data onto the data lines. A reset gate has a circuit depth of only 1, and does this job in the uncompute state. Therefore, replacing an MCX gate with a reset gate further optimizes the circuit parameters.
%, which helped reduce the number of controls needed to trigger the CNOT control line. 
We calculate the gate count and circuit depth, and compare them with the corresponding values for the naive implementation of the QROM circuit to further quantify the benefit of pre-decoding in QROM circuit.

\section{Results and Limitations}
\subsection{Results} 
Since subsets of address lines are used to pre-decode and obtain the pre-decoded signals, multiple such combinations of subsets are possible to offer a tradeoff space among circuit depth, gate count, compilation time and number of extra qubits. For example, a few possible cases for 5 address lines can be,
\begin{enumerate}
    \item Pre-decode 2 address lines and leave 3 address lines undecoded (2(\textbf{P})+3(\textbf{U}))
    \item Pre-decode 3 address lines and leave 2 address lines undecoded (3(\textbf{P})+2(\textbf{U}))
    \item Pre-decode two pairs of 2 address lines and leave the leftover 1 address line undecoded (2(\textbf{P})+2(\textbf{P})+1(\textbf{U}))
    \item Pre-decode 2 address lines and pre-decode rest 3 address lines (2(\textbf{P})+3(\textbf{P}))
    \item Pre-decode 4 address lines together, and leave 1 address line undecoded (4(\textbf{P})+1(\textbf{U}))
    \item Pre-decode all 5 address lines (5(\textbf{P}))
\end{enumerate}

We denote the subset of address lines that are pre-decoded by `\textbf{P}', and the undecoded subset of address lines by `\textbf{U}'. As mentioned previously, various combinations of sets of `P' and `U' subsets of addresses are feasible. We obtained the compilation times, gate counts, and circuit depths for all possible combinations for a particular address width. It is found that the most optimal results are obtained with $\frac{n}{2}$(P)$+\frac{n}{2}$(P) configuration for $n$ address lines in general. An example is provided in Fig. \ref{fig:2p_2p} where 4 address lines are broken down as 2(\textbf{P})+2(\textbf{P}). Fig. \ref{table:naive_results} shows the compilation time, circuit depth, and gate count for the naive and various optimal configurations, including the sawtooth circuit and the two variants of our proposed predecoding circuit. On one hand, it shows drastic reduction in all values due to reduction of the control signals required for the MCX gates in the QROM circuit part present inside right box in Fig. \ref{fig:optimized_qrom}. On the other hand, the number of controls signals increases in the pre-decoding circuit. These are the MCX gates present in the pre-decoding part of optimized circuit shown in left box in Fig. \ref{fig:optimized_qrom}. However, the reduction of control signals of the MCX gates which are present after the pre-decoding circuit are more prominent compared to the increase in pre-decoding circuit. This is because the corresponding drop in gate count after decomposition in the QROM circuit is more than the increase in the gate count after decomposition in the pre-decoding circuit. Thus, the overall gate count reduces, leading to a reduction in circuit depth as well.

\begin{figure}[b]
\includegraphics[width=\linewidth]{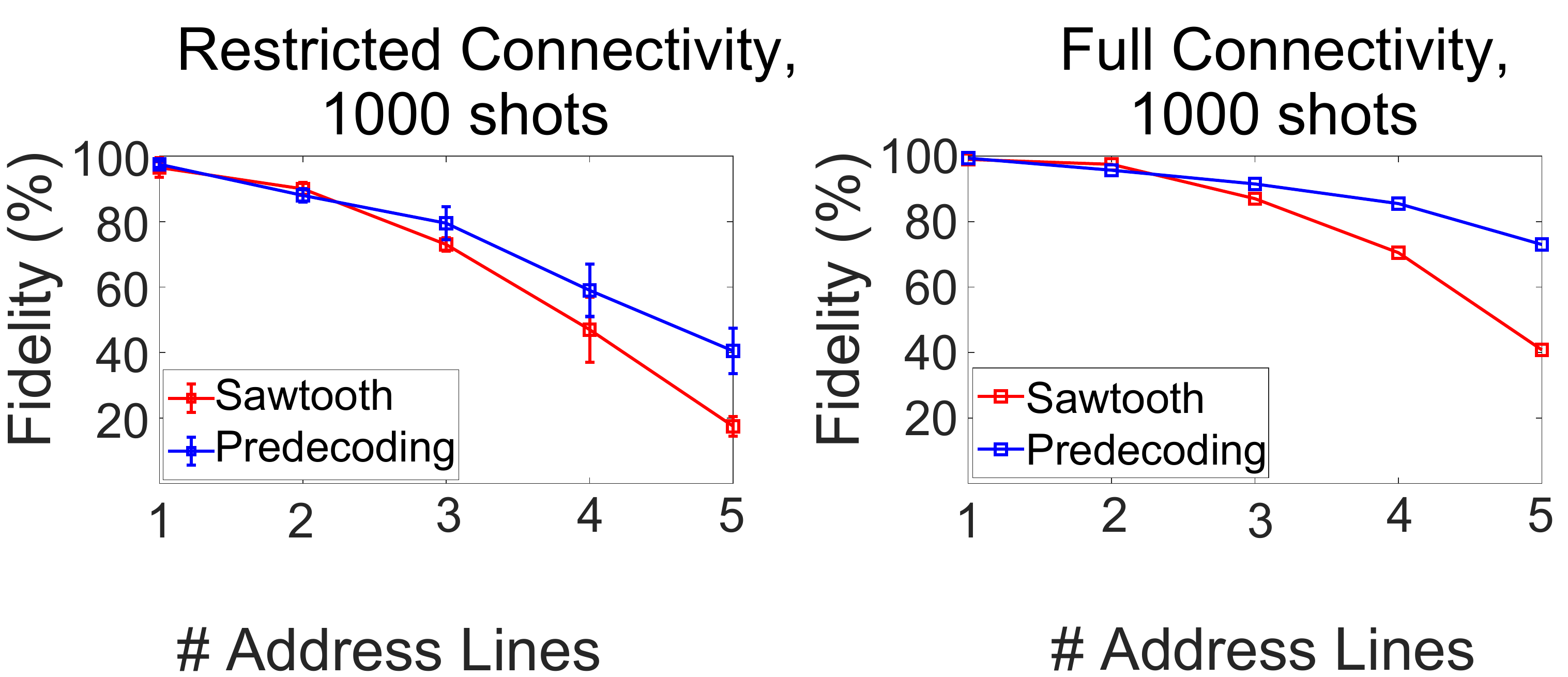}
\caption{Experimental simulation results for 1000 shots for optimized QROM circuits. In one scenario, a restricted qubit connectivity was kept. In the second scenario, fully qubit connectivity was maintained.}
\label{fig:simulation_results}
\end{figure}

Noisy simulations of the optimized QROM circuits are also performed. From the gate count and circuit depth results obtained during compilation, the expected trend is that the fidelity of the predecoding circuit should be higher than the fidelity of the sawtooth circuit at iso- address widths. Also, since the circuit depth increases with the number of address lines, the fidelity should also reduce with increasing number of address lines. For the simulations, two different setups were used. In one setup, restricted qubit connectivity is maintained. The connectivity is given according to the coupling map of IBM Mumbai which is one of IBM's quantum computers running on Falcon processor. In the second setup, full qubit connectivity is kept. For both the setups, a noisy Aer simulator from Qiskit is used, with 0 error rate for single qubit gates, and 0.001 error rate for two qubit gates. This two qubit error rate is approximately one tenth of the actual quantum hardware. The error rate is scaled since otherwise the fidelity values are extremely low with the deep QROM circuits and the expected trend is not clearly visible i.e., output becomes random both with and without optimizations. 
%Thus, the simulations are performed at reduced noise environment.  
The experiments for both setups are run for 1000 shots. Fig. \ref{fig:simulation_results} shows the plots for both the setups for both the optimized circuits. The plots follow the expected trends as mentioned above. For restricted connectivity, the fidelity drop in sawtooth circuit (98\%-19\%) was more than predecoding circuit (98\%-44\%). For full connectivity, the trend was better due to lesser depth (99\%-40.8\% sawtooth, 99\%-73\% predecoding). For the first scenario of restricted connectivity, error bars have to be used because the fidelity values are volatile and fluctuate a lot. The reason for this fluctuation is due to restricted qubit connectivity leading to an extra step of swap insertion procedure to adhere to the physical qubit mapping, thereby increasing the circuit depth. With this increased circuit depth, there is be more fidelity degradation.

\begin{figure}[b]
    \centering
    \includegraphics[width=0.5\linewidth]{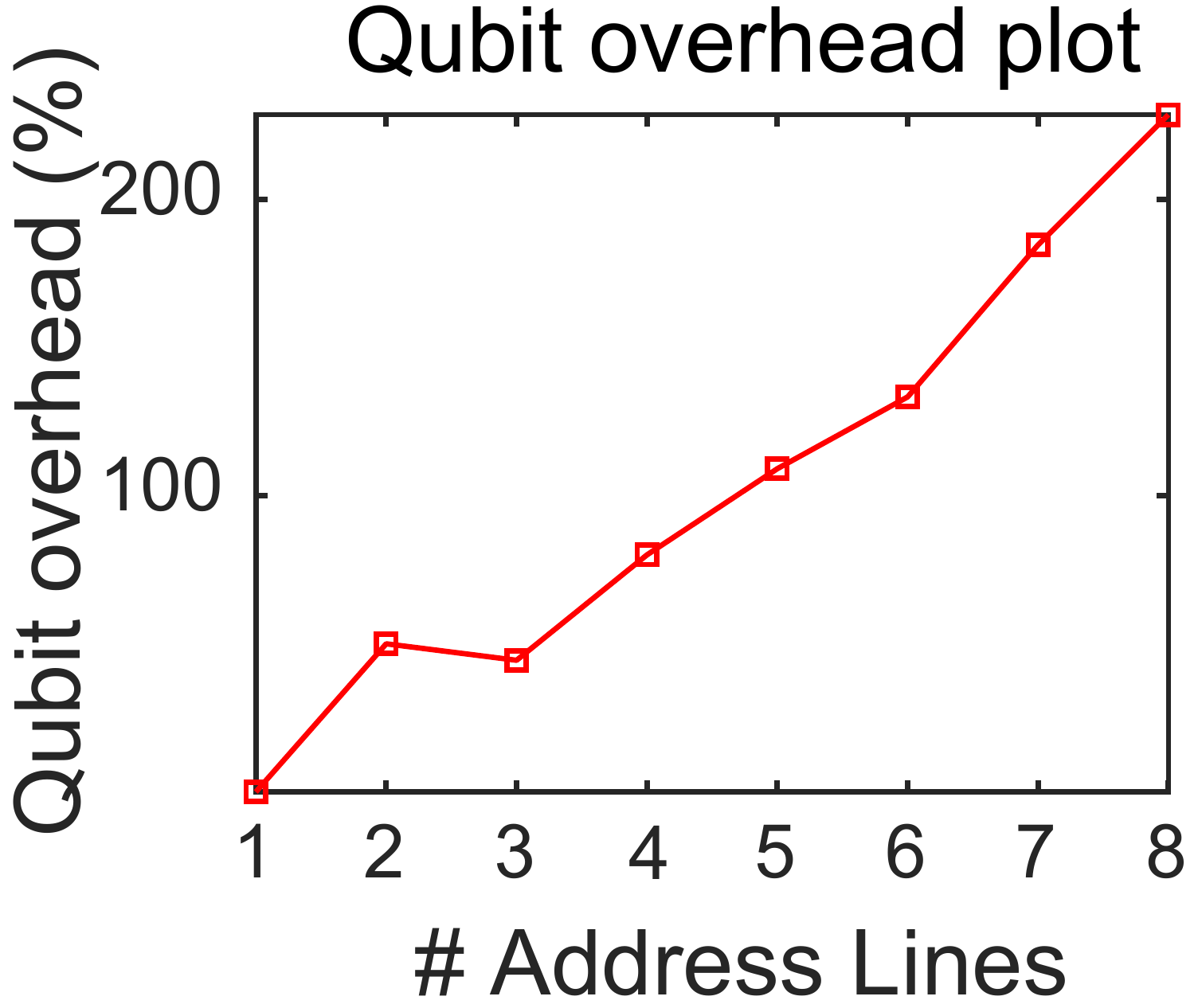}
    \caption{Qubit overhead in optimized QROM circuit for different address lines for the optimal configuration of $\frac{n}{2}$(P)$+\frac{n}{2}$(P). As the number of controls of MCX gates in the pre-decoding configuration increases, the qubit overhead also increases.}
    \label{table:qubit_overhead}
\end{figure}

\subsection{Limitations}
One should recall that the reduction in circuit depth, gate count and compilation time is at the cost of circuit width i.e., $2^m$ extra ancilla qubits for every $m$ subset of pre-decoded address lines.
%, $2^m$ signals are needed, which means tha. 
We compare the total number of qubits required for both naive implementation and the optimized QROM circuit. The number of qubits required for naive QROM circuit can be calculated as follows: 1 qubit for read control line, $n$ qubits for $n$ address lines, 1 qubit for CNOT control line and $d$ qubits for $d$ data lines. Therefore, the total number of qubits will be $1+n+1+d=n+d+2$. In this case, we are keeping $d$ at a constant value of 4. Therefore, the total number of qubits required will be $n+4+2=n+6$. Using this as the reference, we calculate the qubit overhead of the optimized QROM circuit for the optimal configuration of $\frac{n}{2}$(P)$+\frac{n}{2}$(P). The results have been plotted in Fig. \ref{table:qubit_overhead}. The general trend observed is that the number of ancilla qubits required in the pre-decoding circuit increases ($\frac{n}{2}$) with the number of control signals since more number of signals will be pre-decoded. The qubit overhead is therefore more in such cases. There are few minor deviations from this trend. For example, the qubit overhead at 2 qubits is 50\%, while that at 3 qubits is 44.44\%. This is because the extra qubits needed is same (i.e., 4) while the number of naive qubits needed overall increases from 8 to 9 reducing the \% qubit overhead. %But otherwise, it's a generally increasing trend. 

From the results obtained, we note a as high as around 75X reduction in the circuit depth and gate count, and 85X in the compilation time at the cost of $\approx 2.3$X extra qubits for 8 address lines. This improvement will further increase as the number of address lines increase. If the qubit overhead is large, one can further break down the optimal configuration into $\frac{n}{4}$(P)$+\frac{n}{4}$(P)$+\frac{n}{4}$(P)$+\frac{n}{4}$(P) to reduce the overhead at the expense of increased circuit depth and gate count.

\begin{figure}[t]
    \centering
    \includegraphics[width=\linewidth]{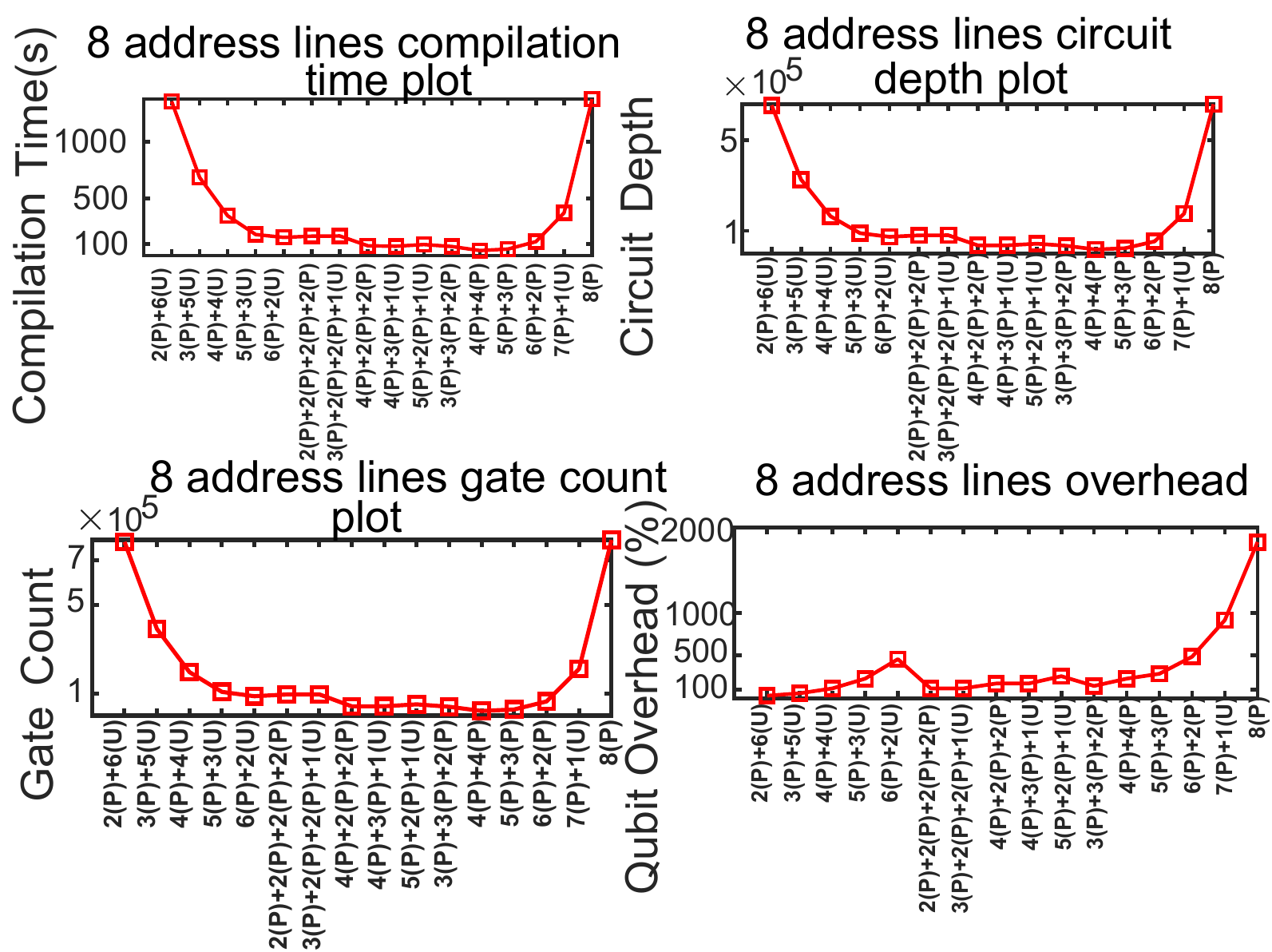}
    \caption{Compilation time, gate count, circuit depth, and qubit overhead plots for different configurations of 8 address lines.}
    \label{fig:8_config}
\end{figure}

To get a deeper understanding of the behavior of different configurations of QROM circuits, we performed further analysis of QROM circuits at different configurations of the same number of address lines. Fig. \ref{fig:8_config} shows the compilation time, gate count, circuit depth, and qubit overhead plots for different configurations of 8 address lines. As mentioned previously, we found that the optimal values are obtained at $\frac{n}{2}$(\textbf{P})$+\frac{n}{2}$(\textbf{P})$=4$(\textbf{P})$+4$(\textbf{P}) configuration. This however, comes at the cost of $2^4+2^4=32$ extra ancilla qubits required in pre-decoding. Another observation is that the values go high when either there are lot of undecoded lines, or when a lot of address lines are pre-decoded together into a single control. Therefore, it is prudent to have a balance of equally pre-decoded address lines and less undecoded lines to keep both qubit overhead and rest of the values as small as possible.

In terms of experiments, it is possible to simulate up to 5 address lines in noisy simulation at higher computational power demand due to increased circuit depth (a limitation). Moreover, as mentioned above, the simulations are performed in reduced noise environment. The noisy simulations also assume full qubit connectivity, which is not the case for real quantum hardware. As a result, while implementing this circuit on real quantum computers, error correction methods like the ones shown in \cite{cory1998experimental, chiaverini2004realization, terhal2015quantum} are required to mitigate the fidelity degradation and get more accurate measurement outputs. Nevertheless introduction of a memory element such as the one proposed herein could revolutionize practical quantum computing as it does not exist currently.

\section{Discussion}
The proposed optimization
%s to the naive implementation of the unary iteration subcircuit of a QROM circuit to 
reduces the gate count, circuit depth and compilation time at increased circuit width. This approach is still practical as qubit counts are growing over the years asymmetrically than the quantum volume. %Companies that are leading the race for quantum supremacy have laid out roadmaps to realize larger quantum computers with lot of qubits. 
IBM's largest quantum computer has 127 qubits (\cite{ibmqc}) with plans to build quantum computers with greater than 1000 qubits by 2023 (\cite{ibm2023}). Google has similar plans for scaling their quantum technology to more than 100,000 qubits (\cite{quantumai}). 
%D-Wave Systems have plans to implement Advantage $2^{TM}$ QPU, which has 20 qubit connectivity and will accommodate 7000+ qubits by 2023-24 (\cite{dwavesys}). 
Therefore, sacrificing qubits to improve the fidelity of computation is a viable direction.

One may argue from the experimental results that the QROM circuits are not yet very practical due to fidelity degradation caused by noise in NISQ computers. While this is indeed somewhat true for current NISQ era computers, this issue will eventually die down as improvements are made in quantum computers in general. According to \cite{preskill2018quantum}, error rates of quantum hardware in the future will reduce significantly, this in turn indicates larger circuits with bigger depths and gate counts and the proposed quantum ROM architecture will run with much higher fidelity, completely eliminating a potential practicality counter argument. Along with this, emerging applications do have a need for quantum memories. These will soon increase the demand for proposed quantum memory to be readily available as a building block. Our work targets this anticipated demand in a timely manner.

\section{Conclusion}

Quantum memory is an important element that can potentially accelerate applications such as, quantum machine learning. Conventional QROM circuits suffer from high depth, large gate count and higher compilation time for wider address sizes. We presented a pre-decoding and reset technique to improve the performance of QROM circuits. We noted reduction in circuit depth and gate count as high as 75X and compilation time by 85X at the cost of a maximum of 2.28X qubit overhead. A lesser fidelity drop was also observed in the predecoding circuit compared to the sawtooth circuit.

\begin{acks}
We acknowledge the use of IBM Quantum Services for this work. The views expressed are those of the authors, and do not reflect the official policy or position of IBM or the IBM Quantum team.
\end{acks}
\vspace{-0.1cm}
\bibliographystyle{ACM-Reference-Format}
\bibliography{acmart}

% \appendix









\end{document}